\documentclass{article}

\usepackage{authblk}
\usepackage{graphicx}
\usepackage{hyperref}
\usepackage{acronym}
\usepackage{amsmath}
\usepackage{mathtools}
\usepackage{siunitx}
\usepackage{cleveref}
\usepackage{fullpage}

\newacro{ABF}{artificial bacterial flagellum}
\newacroplural{ABF}{artificial bacterial flagella}
\newacro{DPD}{dissipative particle dynamics}
\newacro{SDE}{stochastic differential equation}
\newacro{SDF}{signed distance function}
\newacro{SFS}{stress-free shape}
\newacro{RBC}{red blood cell}
\newacro{RL}{reinforcement learning}

\newcommand{\Ht}{\mathrm{Ht}}

\makeatletter
\newcommand\etal{et al\@ifnextchar.{}{.\ }}
\makeatother

\begin{document}

\title{Optimal navigation of magnetic artificial microswimmers in blood capillaries with deep reinforcement learning}

\author[1]{Lucas Amoudruz}
\author[1]{Sergey Litvinov}
\author[1,$\ast$]{Petros Koumoutsakos}
\affil[1]{School of Engineering and Applied Sciences, Harvard University, Cambridge, MA 02138, United States.}
\affil[$\ast$]{corresponding author: petros@seas.harvard.edu}

\maketitle

\begin{abstract}
  Biomedical applications such as targeted drug delivery, microsurgery, and sensing rely on reaching precise areas within the body in a minimally invasive way.
  Artificial bacterial flagella (ABFs) have emerged as potential tools for this task by navigating through the circulatory system with the help of external magnetic fields.
  While their swimming characteristics are well understood in simple settings, their controlled navigation through realistic capillary networks remains a significant challenge due to the complexity of blood flow and the high computational cost of detailed simulations.
  We address this challenge by conducting numerical simulations of ABFs in retinal capillaries, propelled by an external magnetic field.
  The simulations are based on a validated blood model that predicts the dynamics of individual red blood cells and their hydrodynamic interactions with ABFs.
  The magnetic field follows a control policy that brings the ABF to a prescribed target.
  The control policy is learned with an actor-critic, off-policy reinforcement learning algorithm coupled with a reduced-order model of the system.
  We show that the same policy robustly guides the ABF to a prescribed target in both the reduced-order model and the fine-grained blood simulations.
  This approach is suitable for designing robust control policies for personalized medicine at moderate computational cost.
\end{abstract}

\section{Introduction}

Targeted drug delivery, microsurgery, and microsensing represent important areas of research aimed at revolutionizing precision medicine~\cite{soto2020medical,manzari2021targeted}.
These challenging tasks require precise access to target areas in the body, which are often difficult to reach in a non-invasive way.
During the past decade, artificial microswimmers have emerged as potent candidates to navigate specific regions of the human body through the circulatory system or tissues~\cite{bunea2020}.
A particular form of artificial microswimmers, known as \acp{ABF}, features a cork-screw-shaped body propelled by rotating magnetic fields~\cite{dreyfus2005}.
\Acp{ABF} have been used in various applications, including navigation through ex-vivo bovine eye tissues~\cite{yu2019}, assisting spermatozoa in reaching ovocytes~\cite{medina2016}, enhancing nanoparticle transport~\cite{schuerle2019}, and carrying drugs for cancer therapy~\cite{wang2019}.

The swimming properties of \acp{ABF} are well understood in the idealized scenario of unbounded, viscous Newtonian fluids~\cite{zhang2009artificial,zhang2009characterizing}, near walls~\cite{pal2022fluid}, and near other swimmers~\cite{dey2022oscillatory}.
Venugopalan \etal demonstrated that \acp{ABF} can propel in low-concentrated blood suspensions~\cite{venugopalan2014}.
Alapan \etal have guided microrollers in physiologically relevant blood concentrations with a rotating magnetic field~\cite{alapan2020multifunctional}.
Qi \etal have conducted numerical simulations of these rollers through blood in straight pipes~\cite{qi2021quantitative}.
However, the swimming characteristics of \acp{ABF} within the bloodstream have not been thoroughly investigated.
The control of microswimmers for path planning has been the subject of recent advances with the emergence of \ac{RL}.
Mui\~nos-Landin \etal used \ac{RL} in an experimental setup to control a microswimmer towards a target, without background flow~\cite{muinos2021reinforcement}.
Similarly, several studies applied \ac{RL} to guide microswimmers~\cite{mo2023challenges} through background flows~\cite{colabrese2017flow,colabrese2018smart,biferale2019,alageshan2020,borra2022reinforcement} and perform independent control of multiple \acp{ABF}~\cite{amoudruz2022,karnakov2025optimal} to reach a target or follow a prescribed path~\cite{liu2025reinforcement}.
Yang \etal achieved path-planning of point particles between obstacles represented by \acp{RBC}~\cite{yang2022hierarchical}.
However, the numerical models used in these studies do not reflect the complex flow patterns present in capillaries, the finite size of the swimmer, and the presence of deformable blood cells, which are known to strongly affect blood flow~\cite{amoudruz2024volume}.

In this work, we model and simulate an \ac{ABF} that evolves in a retinal capillary network.
In particular, we learn a control policy that guides the \ac{ABF} to a prescribed target by imposing an external magnetic field.
The geometric representation of the environment is reconstructed from a fundus image of a retinal capillary network~\cite{ghassemi2015}.
The simulations include an accurate \ac{RBC} model that was extensively validated on experimental data~\cite{amoudruz2023a}, coupled with the \ac{DPD} method to resolve the hydrodynamics at the microscale~\cite{groot1997}.
The control policy is learned with the actor-critic, off policy \ac{RL} algorithm V-RACER~\cite{novati2019a}.
To substantially reduce the computational cost associated with expensive simulations of the environment, the \ac{RL} agent is trained on a reduced order model of fine-grained blood simulations.
The approach of training the \ac{RL} agent on a reduced order model has been used in the context of robotics to reduce the number of experiments or expensive simulations~\cite{marco2017virtual,peng2018sim}.
We demonstrate that the policy is robust to noise and transferable to the fine-grained and expensive simulation environment.
These results indicate that \ac{RL} is a promising approach to design control policies with the level of robustness required in biomedical applications.
Furthermore, this work demonstrates that \acp{ABF} can evolve within the intricate geometry and flow fields of the retinal capillaries to reach precise locations with minimal invasiveness.

\section{Artificial microswimmer within blood flow}

\begin{figure*}
  \centering
  \includegraphics[width=0.9\textwidth,trim={0 3cm 0 3cm},clip]{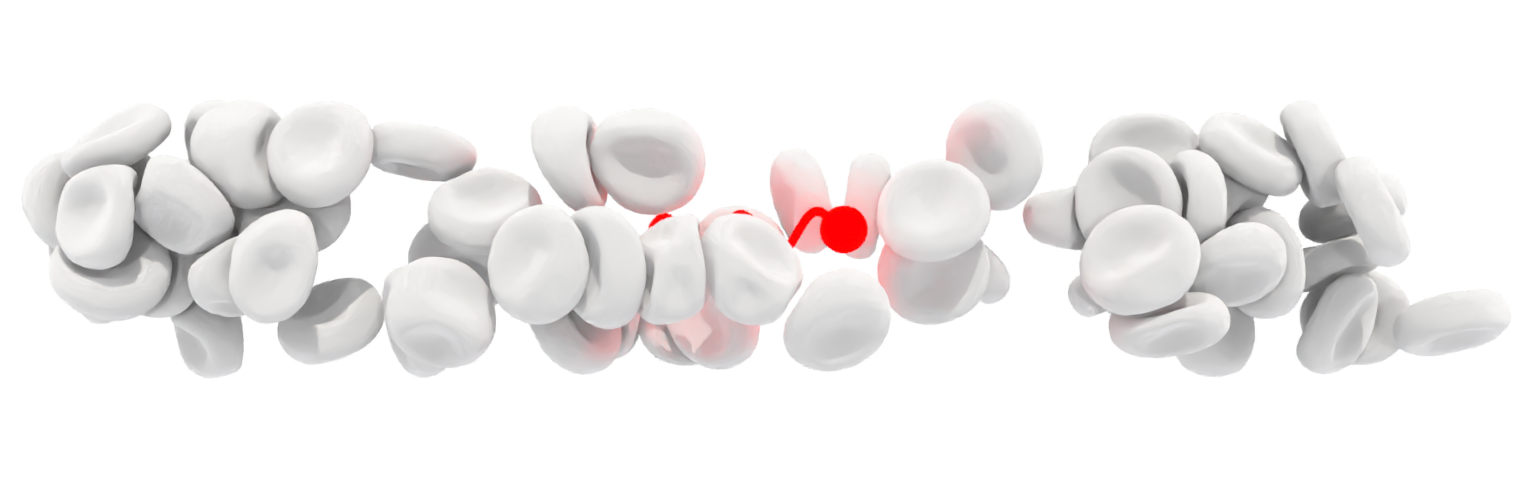}
  \caption{
    Simulation snapshot of an ABF inside a periodic tube filled with blood.
    Tube boundaries not shown for visualization purpose.
  }
  \label{fig:snapshot:tube}
\end{figure*}

The simulations consist of an \ac{ABF} evolving through a network of capillaries filled with blood at a hematocrit $\Ht=25\%$.
The \ac{ABF} has a radius of $\SI{2}{\micro\meter}$ and a length of $\SI{18.37}{\micro\meter}$, within the range of the \acp{ABF} presented in the literature~\cite{zhang2009artificial,zhang2009characterizing}.
Blood is modeled as \ac{RBC} membranes suspended in plasma and enclosing the cytosol, a fluid 5 times more viscous than plasma~\cite{dasanna2021importance}.
Both plasma and cytosol are viscous Newtonian fluids and are modeled with the \ac{DPD} method~\cite{groot1997}.
\Ac{RBC} membranes evolve according to bending forces, and shear forces that are zero when the membrane takes the shape of the stress-free state of the \ac{RBC} cytoskeleton~\cite{julicher1996morphology,lim2008,amoudruz2023a}.
An example of simulation of an \ac{ABF} swimming through a straight tube filled with blood is shown in \cref{fig:snapshot:tube}.

The bending energy of the membrane is given by
\begin{equation} \label{eq:energy:bending}
  U_{bending} = 2 \kappa_b \oint{H^2dA},
\end{equation}
where the integral is taken over the membrane surface, $\kappa_b$ is the bending modulus, and $H$ is the local mean curvature.
The energy related to the membrane deformation with respect to the \ac{SFS} is given by
\begin{multline} \label{eq:energy:inplane}
  U_{in-plane} = \frac {K_\alpha}{2} \oint {\left( \alpha^2 + a_3 \alpha^3 + a_4 \alpha^4 \right) dA_0} \\
  + \mu \oint{ \left( \beta + b_1 \alpha \beta + b_2 \beta^2 \right) dA_0},
\end{multline}
where the integral is taken over the surface of the \ac{SFS}, $\alpha$ and $\beta$ are the local dilation and shear strain invariants of the membrane, respectively, $K_\alpha$ is the dilation elastic modulus, $\mu$ is the shear elastic modulus and the coefficients $a_3$, $a_4$, $b_1$ and $b_2$ are parameters that control the non-linearity of the membrane elasticity for large deformations~\cite{lim2008}.

In addition, the area and volume of the \ac{RBC} membranes are restricted by penalization energies~\cite{fedosov2010pHD},
\begin{equation}
  U_{area} = k_A \frac{\left(A - A_0\right)^2}{A_0}, \quad
  U_{volume} = k_V \frac{\left(V - V_0\right)^2}{V_0},
\end{equation}
where $A_0$ and $V_0$ are the target area and volume of the cell, and $A$ and $V$ are the area and volume of the cell, respectively.
Furthermore, $k_A$ and $k_V$ are coefficients that were chosen empirically to enforce small variations of the area and volume over time, within $1\%$.

Each membrane surface is discretized into a triangle mesh composed of $N_v$ vertices with positions $\mathbf{r}_i$, velocities $\mathbf{v}_i$ and mass $m$, $i=1,2,\dots,N_v$, evolving through time according to Newton's law of motion.
The forces acting on each particle are computed by taking the negative gradient of the discretized energies listed above with respect to particle positions.
Finally, the membrane viscosity is modeled through pairwise forces between particles that share an edge,
\begin{equation}
  \mathbf{f}_{ij}^{visc} = - \frac{4 \eta_m}{\sqrt{3}} \left(\mathbf{v}_{ij} \cdot \mathbf{e}_{ij} \right) \mathbf{e}_{ij},
\end{equation}
where $\eta_m$ is the membrane viscosity, $\mathbf{v}_{ij} = \mathbf{v}_i - \mathbf{v}_j$ and $\mathbf{e}_{ij}$ is the unit vector between $\mathbf{r}_i$ and $\mathbf{r}_j$.
Details about the fluid-structure interactions and model parameters are described in ref~\cite{amoudruz2022phd} with extensive validation against experimental data~\cite{amoudruz2023a}.

The \ac{ABF} is represented as a set of frozen particles and a surface moving as a rigid body.
The surrounding solvent particles interact with the \ac{ABF} particles through \ac{DPD} forces and are bounced-back from the surface.
The magnetic moment $\mathbf{m}$ of the \ac{ABF} remains constant in the reference frame of the swimmer, perpendicular to its principal axis.
The \ac{ABF} is immersed in an external, uniform magnetic field $\mathbf{B}$, creating a magnetic torque $\mathbf{T} = \mathbf{m} \times \mathbf{B}$.
This torque, combined with the cork-screw shape of the \ac{ABF}'s tail, causes the \ac{ABF} to propel along its main axis~\cite{zhang2009artificial}.
The external magnetic field varies over time and will be the controlled quantity used to stir the \ac{ABF} towards its target.

The boundaries of the capillaries were generated from a human retinal vasculature fundus image in Ghassemi et al.~\cite{ghassemi2015} (see Supplementary Material).
The walls are formed by a layer of frozen \ac{DPD} particles, of width larger than the interaction cutoff.
Furthermore, the \ac{DPD} particles are bounced-back from the wall surface.
The parameters of the \ac{DPD} forces and interactions between every objects of the simulations are detailed in Amoudruz~\cite{amoudruz2022phd}.
To drive the flow, an external body force is applied to the \ac{DPD} particles, following the approach described in Yazdani \etal~\cite{yazdani2016flow}.
The body force is derived from the pressure gradient obtained by solving the Stokes equation in the same geometry.
The flow is solved with \textit{Aphros}~\cite{karnakov2020aphros}, assuming a Newtonian fluid, no-slip boundary conditions at the geometry boundaries and a constant velocity at the inlet.
In the fine-grained blood simulations, the body forces were scaled to match the Reynolds number at the inlet of the domain.
Blood simulations are performed with \textit{Mirheo}~\cite{alexeev2020mirheo}, a high-performance package for microfluidic simulations on multi-GPU architectures.

\begin{figure}
  \centering
  \includegraphics{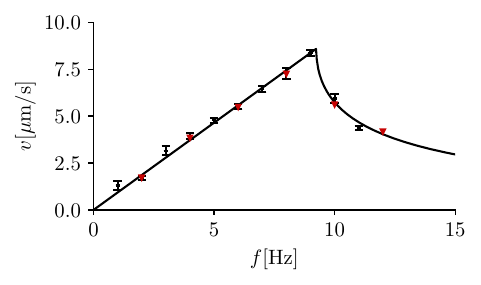}
  \caption{Swimming speed of the ABF against the rotation frequency of the external magnetic field.
  Triangles are obtained with DPD simulations, crosses are from experiments from Mhanna et al.~\cite{mhanna2014artificial}, solid line is a fit to the experimental data with an ODE model~\cite{schamel2013chiral,vach2013selecting} (see Supplementary Material).}
  \label{fig:abf:validation}
\end{figure}

\Cref{fig:abf:validation} shows the swimming speed of a single \ac{ABF} in a viscous fluid against the rotation frequency of the external magnetic field.
The \ac{ABF} used for this validation case is a helix of diameter $\SI{5.294}{\micro\meter}$ and length $\SI{15.75}{\micro\meter}$ with 3 turns and an ellipsoidal cross section with main diameters $\SI{1.192}{\micro\meter}$ and $\SI{2.231}{\micro\meter}$.
This shape was reproduced to match that of the \ac{ABF} in experiments from Mhanna et al.~\cite{mhanna2014artificial}.
The magnetic moment of the swimmer was set to $\SI{1e-14}{\newton\meter\per\tesla}$, tuned to match the experimental data, and the magnetic field magnitude to $\SI{3}{\milli\tesla}$.
Simulations are in good agreement with the experiments by Mahnna et al.~\cite{mhanna2014artificial}, validating the \ac{DPD} method to simulate swimming \acp{ABF} in fluids at low Reynolds numbers.

In this work, the \ac{ABF} has a helical tail and a spherical head, designed to prevent the \ac{ABF} from getting stuck in \acp{RBC}.
To increase the swimming speed of the \ac{ABF}, both the magnetic field frequency and the magnetic moment of the \ac{ABF} were increased to $f = \SI{1}{\kilo\hertz}$ and $\SI{1e-11}{\newton\meter\per\tesla}$. respectively.
These adjustments enable the \ac{ABF} to reach swimming speeds comparable to the inlet flow velocity $U_{in} = \SI{1}{\milli\meter\per\second}$.
Such a high magnetic moment can be achieved by using a magnetic head composed, for example, of nickel, which has a magnetization of $M_\mathrm{Ni} = \SI{4.9e5}{\ampere\per\meter}$.
For a spherical head of radius $R = \SI{2}{\micro\meter}$, this yields a magnetic moment of $m = M_\mathrm{Ni} 4 \pi R^3 / 3 \approx \SI{1.6e-11}{\newton\meter\per\tesla}$, demonstrating that the values used in simulations are within experimentally achievable values.
Importantly, even at this increased rotation frequency $f=\SI{1}{\kilo\hertz}$, the Reynolds number remains low, $\mathrm{Re} = R^2 f / \nu \approx 0.004$, justifying the use of the Stokes equation.
Therefore, the validation shown in \cref{fig:abf:validation} indicates that the \ac{DPD} method remains appropriate to predict the evolution of \acp{ABF} rotating with an angular frequency of $f = \SI{1}{\kilo\hertz}$.

\section{Learning a RL policy with a reduced order model}

In this section, we describe the procedure used to learn, with \ac{RL}, a control policy for guiding the \ac{ABF} through the retinal capillary network.
The goal is to bring the \ac{ABF} from the network entry point to a specified target location.
This is achieved by varying the external magnetic field over time.
\Ac{RL} algorithms typically require tens of thousands to millions of episodes to converge.
However, each fine-grained simulation of the \ac{ABF} in the capillary network has a high computational cost of 64 P100 GPUs over 24 hours.
Therefore, it is intractable to train the \ac{RL} agent directly on the high-fidelity model.
Instead, the \ac{RL} agent is coupled with a reduced-order model during training.

The reduced order model represents the \ac{ABF} as a self-propelling point particle advected by a background velocity field $\mathbf{u}$.
The background velocity field is obtained by solving the Stokes equation within the geometry of the retinal network.
The Stokes equation was computed with the grid-based solver \textit{Aphros}.
In addition, collisions between the \ac{ABF} and surrounding \acp{RBC} are modeled by a stochastic term, resulting in the \ac{SDE}
\begin{equation} \label{eq:sde:rl}
  \dot{\mathbf{x}} = \mathbf{u}(\mathbf{x}) + U \mathbf{p} + \sqrt{D} \boldsymbol\xi,
\end{equation}
where $\mathbf{p}$ and $U$ are the direction and magnitude of the \ac{ABF} self-propelling velocity, respectively.
$D$ is a diffusion coefficient and $\boldsymbol\xi$ is a Gaussian white noise vector.
In addition, the particle bounces back from the wall boundaries.
The parameters $D$ and $U$ were calibrated from high-fidelity blood flow simulations of a single \ac{ABF} in a straight pipe of radius typically found in capillaries (see Supplementary Material).
We neglect the rotation dynamics of the \ac{ABF}, as the reorientation time scale $1/f = \SI{e-3}{\second}$ is 20 times faster than the typical time scale associated with flow velocity gradients, $2R_{in}/U_{in} = \SI{2e-2}{\second}$.

The system is advanced in time with a piecewise constant action $\mathbf{p}$, updated every $\Delta t$ units of time.
The direction $\mathbf{p}$ is computed from the \ac{ABF}'s position $\mathbf{x}$ through the control policy, $\mathbf{p} = \hat{\mathbf{p}} / |\hat{\mathbf{p}}|$, $\hat{\mathbf{p}} = \pi(\mathbf{x})$.
Each episode ends if the simulation time exceeds a maximum time $T_\text{max}$ or if the \ac{ABF} reaches the target within a distance $\delta$.
The initial position of the \ac{ABF} at the beginning of each episode is sampled from a set of positions that are along the downstream vessels of the target (see Supplementary Material).
The reward at step $t$ is expressed as
\begin{equation} \label{eq:rl:reward}
  r_t = -C\Delta t + \|\mathbf{x}_{t-1} - \mathbf{x}_\text{target}\| - \|\mathbf{x}_{t} - \mathbf{x}_\text{target}\|,
\end{equation}
where $C>0$ is a constant and $\mathbf{x}_\text{target}$ is the target position.
The first term in \cref{eq:rl:reward} penalizes long trajectories, while the second term is a reward shaping that is positive when the \ac{ABF} progresses towards its target~\cite{ng1999}.

We train the policy using V-RACER~\cite{novati2019a}, an actor-critic off-policy \ac{RL} algorithm that was applied to falling objects~\cite{novati2019controlled}, magnetic microswimmers~\cite{amoudruz2022} and self-propelling fish~\cite{verma2018,mandralis2021learning}.
The algorithm is implemented in \textit{Korali}~\cite{martin2022korali}.
Each episode consists of about 500 experiences, and we train the agent over 10\,000 episodes.

\begin{figure}
  \centering
  \includegraphics[width=0.5\columnwidth]{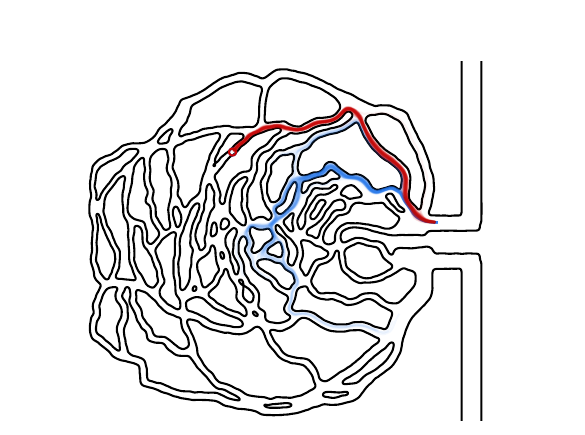}
  \caption{
    Trajectories of passive tracers (blue) and controlled swimmers (red) obtained from 100 random seeds, with the reduced order model and $D=D_\mathrm{sim}$.
    The circle represents the target.
    The inlet and outlet have a diameter of $\SI{20}{\micro\meter}$.
  }
  \label{fig:rl:trajectories}
\end{figure}

\begin{figure}
  \centering
  \includegraphics{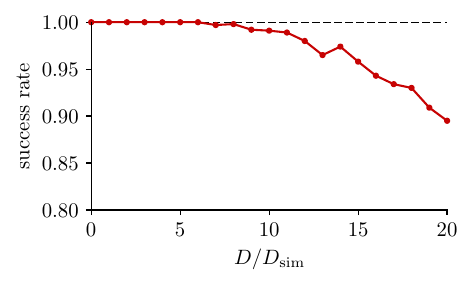}
  \caption{
    Success rate of the control policy against the noise level $D$ relative to that measured from the fine-grained simulations $D_\mathrm{sim}$.
  }
  \label{fig:rl:robustness}
\end{figure}

To test the policy, we generate 100 trajectories with \cref{eq:sde:rl}, both with the trained policy and for passive tracers ($U=0$).
These trajectories are shown in \cref{fig:rl:trajectories}.
Passive tracers are much more sensitive to noise than controlled swimmers, and their trajectories end at various locations in the geometry.
Passive tracers do not reach the target, as opposed to swimmers that follow the \ac{RL} policy.

The robustness of the control policy is assessed by varying the magnitude of noise $D$ and measuring the success rate, defined as the number of times the particle reaches the target out of 1000 trials divided by the number of trials.
\Cref{fig:rl:robustness} demonstrates that the control policy brings the swimmer to its target more than 98\% of the time when $D < 10 D_\mathrm{sim}$, and $100\%$ of the time when the noise is comparable to that measured from the fine-grained simulations, $D_\mathrm{sim}$.
Increasing the noise level further results in lower success rates.
However, the success rate remains above 85\% when $D \leq 20 D_\mathrm{sim}$.
In comparison, passive tracers never reached the target for any values $D\in[0, 20D_\mathrm{sim}]$ in our experiments.

\begin{figure*}
  \centering
  \includegraphics[width=0.45\textwidth]{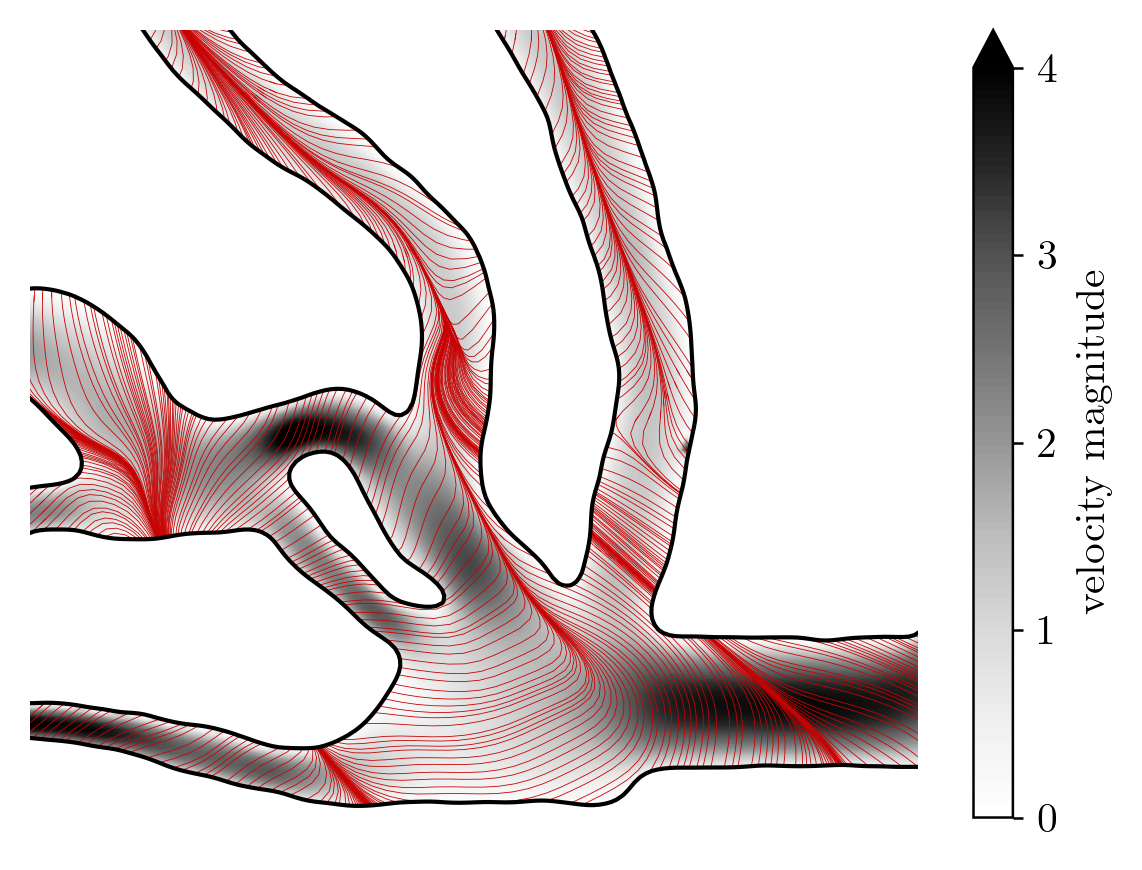}
  \includegraphics[width=0.45\textwidth]{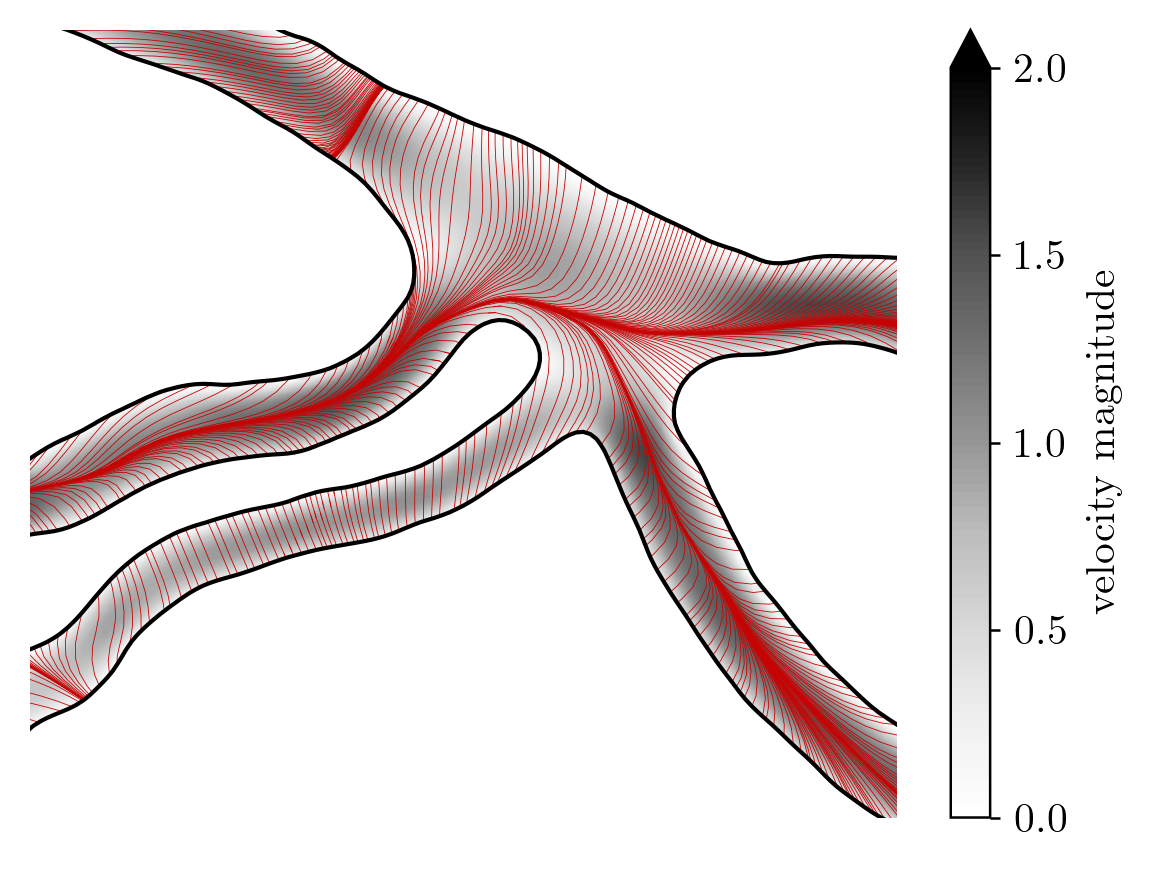}
  \caption{
    Streamlines of the policy at two bifurcations along the optimal path.
    The direction chosen by the agent is parallel to the streamlines.
    The background colors indicate the background velocity magnitude.
  }
  \label{fig:rl:policy}
\end{figure*}

\begin{figure}
  \centering
  \includegraphics[width=0.5\columnwidth]{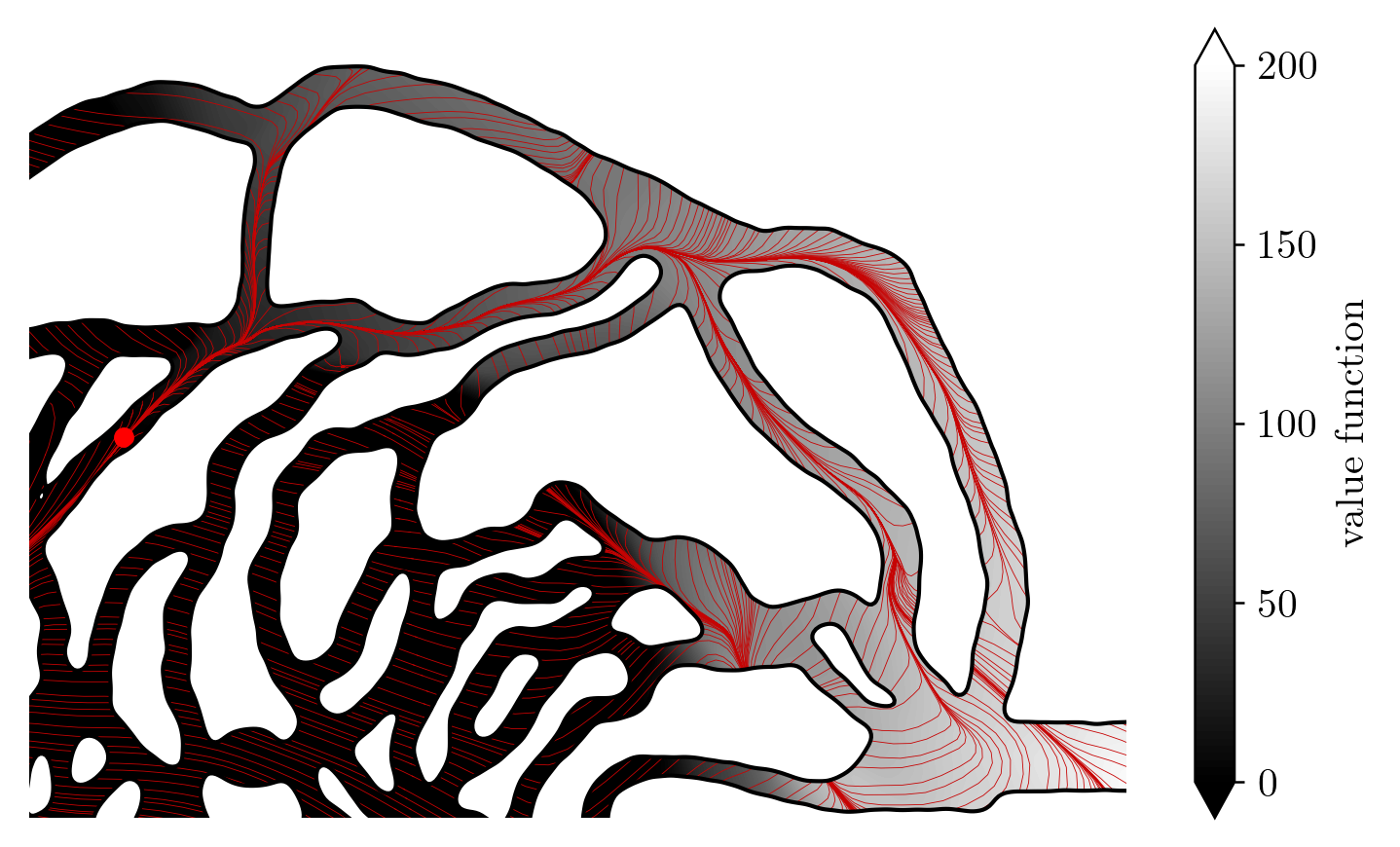}
  \caption{
    Streamlines of the policy within the capillaries (solid lines) and state value function (shades of grey).
    The direction chosen by the agent is parallel to the streamlines.
    The target is represented with the red disk.
  }
  \label{fig:rl:policy:value}
\end{figure}

\Cref{fig:rl:policy} shows the policy at bifurcations.
The streamlines of the directions obtained from the policy, $\mathbf{p} = \hat{\mathbf{p}} / |\hat{\mathbf{p}}|$, $\hat{\mathbf{p}} = \pi(\mathbf{x})$, converge to a line that connects the starting point to the target.
This line seems to favor regions where the background velocity is the largest and aligned with the swimming direction, thus decreasing the overall travel time of the swimmer.
When the swimmer is away from this line, the agent chooses a direction that brings the swimmer back to the line.
This mechanism explains the robustness of the control policy to external perturbations.

\Cref{fig:rl:policy:value} shows the streamlines of the policy, together with the state-value function estimated during learning.
The state-value function takes low values in regions where the swimmer cannot recover due to strong flow fields.
Hence, in these regions, the policy is not meaningful.
In regions away from the optimal path and with higher state-value function, the policy has a similar behavior to the policy along the optimal path: the streamlines converge near the centerline of the vessel, where the velocity is higher.

\section{Control of the ABF in fine-grained blood simulations}

The control policy trained on the reduced-order model is now tested in the fine-grained model.
The policy maps the swimmer's position to the swimming direction $\mathbf{p}$.
However, in the fine-grained simulations, the swimming direction is not set directly and we instead control the magnetic field.
Given that \acp{ABF} align perpendicularly to the plane of the magnetic field's rotation, we adjust the magnetic field as
\begin{equation}
  \mathbf{B}(t) = B R_x(\mathbf{p})
  \begin{pmatrix}
    0 \\
    \cos \omega t \\
    \sin \omega t
  \end{pmatrix},
\end{equation}
where $B$ and $\omega$ are the magnitude and frequency of rotation of the magnetic field, respectively, and $R_x(\mathbf{p})$ is the rotation that transforms the vector $\mathbf{e}_x$ into the swimming direction $\mathbf{p}$ with axis of rotation $\mathbf{e}_x \times \mathbf{p}$.
The direction $\mathbf{p}$ is computed from the policy evaluated at the \ac{ABF}'s center of mass, and the magnetic field is adapted at every time step of the simulation.

\begin{figure*}
  \centering
  \includegraphics[width=0.8\textwidth]{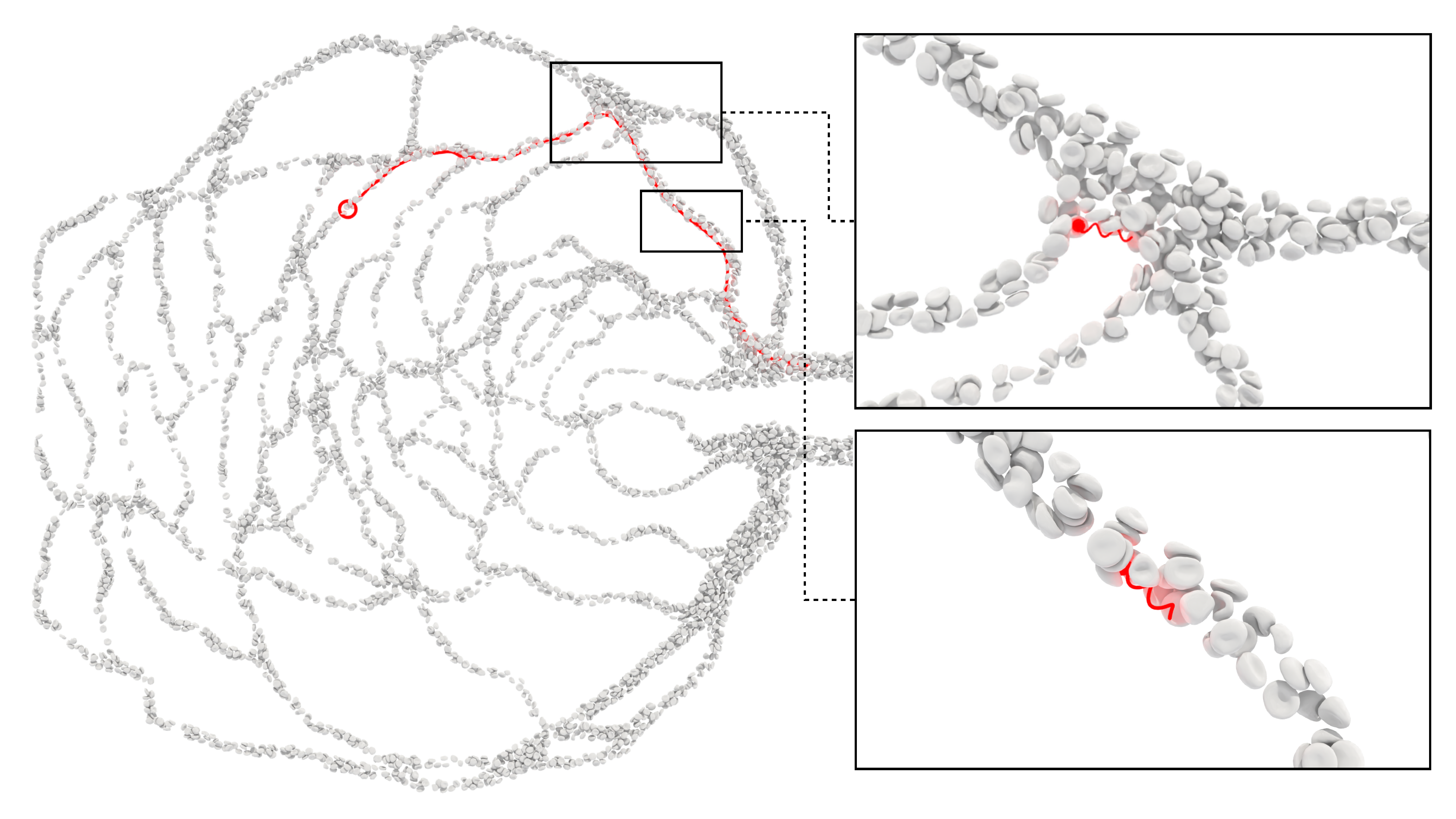}
  \caption{
    Trajectory of the ABF in the blood simulations with feedback control via the RL policy.
    The ABF reaches the target (red circle) and follows a trajectory (red line) similar to that obtained with the reduced order model.
    The zoomed-in views show the ABF swimming in the capillaries.
  }
  \label{fig:rl:full:traj}
\end{figure*}

\Cref{fig:rl:full:traj} (Multimedia available online) shows that the \ac{ABF} successfully reaches the target using the policy that was previously learned from the reduced-order model.
The \ac{ABF} follows a trajectory similar to that observed in the reduced-order model.
Additionally, the \ac{ABF} tends to align with the flow direction and the centerline of the capillaries, suggesting that the trajectory minimizes the travel time.
Away from bifurcations, the policy seems to keep the \ac{ABF} near the center of the capillaries, where the flow is faster.
We observe that the \ac{ABF} remains close to the wall boundaries in the vicinity of bifurcations.
This strategy offers two benefits: the \ac{ABF} is closer to the correct downstream branch, reducing the risk of taking the wrong bifurcation, and the \ac{ABF} has a larger swimming speed relative to the flow, allowing finer control in these critical areas.

We emphasize that the policy was trained on an environment that is different from the fine-grained simulations but has a similar qualitative response.
In the reduced-order model, we assume that the swimmer immediately reorients to the control output, which is not the case in the fine-grained blood model.
Furthermore, the velocity field is different from that obtained from the Stokes solver with Newtonian assumption.
Finally, the swimming speed of the \ac{ABF} with respect to the fluid depends on the local hematocrit, the configuration of the surrounding \acp{RBC}, and the geometry of the capillaries, which we ignore in the reduced order model.
A comparison of the \ac{ABF}'s coordinates against time between the \ac{DPD} simulations and the reduced order model is shown in the Supplementary Material.
In both cases the \ac{ABF} follows the same trajectory and has a similar travel time.
We observe small differences in the swimming velocity, that may be due to the surrounding geometry, wall effects, and local hematocrit.
Despite these differences, the policy learned by the \ac{RL} agent in the reduced-order model is successful in the fine-grained simulations.
This transfer indicates that \ac{RL} is an effective method to learn policies that are robust to changes in the environment.

\section{Outlook}

In this study, we assume that the position of the \ac{ABF} can be estimated in real time.
However, in practice, the localization of micrometer-sized objects in vivo remains a significant challenge.
Promising techniques, such as magnetic particle imaging (MPI)~\cite{bakenecker2019actuation}, ultrasound imaging~\cite{khalil2014magnetic} and fluorescence imaging~\cite{wang2024tracking}, have shown potential for tracking magnetic microswimmers in biological environments, but further advances in spatial resolution are necessary to enable their practical application~\cite{pane2019imaging}.

Furthermore, in this work, we have considered a single \ac{ABF}, while it would be beneficial to control multiple \acp{ABF} to achieve targeted drug delivery or microsurgery.
Controlling multiple \acp{ABF} of different shapes with a uniform magnetic field has been explored in the absence of wall boundaries~\cite{amoudruz2022}, but this approach is currently limited to a few \acp{ABF}.
An other potential approach to control a swarm of \acp{ABF} consists in exploiting the collective behavior of micro-objects through their magnetic interactions~\cite{kokot2018manipulation}, although current research considers ideal setups of microrollers close to flat surfaces and no background flows.

\section{Summary}

We have performed simulations of the evolution of \acp{ABF} in the bloodstream through complex capillaries.
The simulations employ a fine-grained blood model coupled with the \ac{DPD} method to resolve the flow mechanics.
The \ac{ABF} is propelled via an external magnetic field, controlled by a \ac{RL} agent with the aim of guiding the \ac{ABF} towards a prescribed target.
The agent was trained on a reduced order model that was calibrated from fine-grained blood simulations.
This approach is much less computationally demanding than the use of blood simulations in the training phase.
The control policy is robust to external noise in the reduced-order environment and is successful nearly $100\%$ of the time for relatively large perturbations.
Interestingly, the same control policy is also successful in fine-grained blood simulations.

These results demonstrate that control policies trained on reduced-order models are potent candidates for control in fine-grained simulations and possibly in real conditions, while requiring moderate computational resources.
Furthermore, this approach enables the design of personalized control strategies based on patient-specific data, as we have shown in this case by using the reconstruction of capillary networks generated from fundus images.
This method can be generalized to the navigation of microswimmers for targeted drug delivery and microsurgery at precise locations, potentially becoming a critical tool for personalized medicine.

\section*{Supplementary Material}

The Supplementary Material provides details about the generation of the geometry file, a description about the calibration of the reduced-order model from DPD simulations, the parameters of the reinforcement learning algorithm, a comparison between the models, and analysis of the motion of the \ac{ABF} in the \ac{DPD} simulations.

\section*{Acknowledgments}
We acknowledge the computational resources granted by the Swiss National Supercomputing Center (CSCS) under the project ID ``s1160''.

\section*{Conflicts of Interest}
The authors have no conflicts to disclose.

\bibliographystyle{unsrt}
\bibliography{refs}

\end{document}


\maketitle

\section{Generation of the geometry}

\begin{figure}
  \centering
  \includegraphics{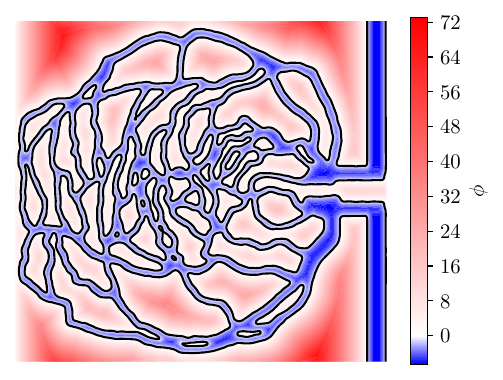}
  \caption{
    Slice $z=12$ of the SDF $\phi(\mathbf{x})$ of the capillaries (grey scale) and corresponding zero-isoline (solid line).
    The SDF is negative inside the capillaries and positive outside.
    This implicit representation provides a convenient way to check whether a position is inside or outside the capillaries.
  }
  \label{fig:sdf}
\end{figure}

The geometric representation of the capillaries is generated from an explicit representation (triangle mesh) obtained from ref.~\cite{ghassemi2015}.
The triangle mesh was smoothed with a Laplacian filter and simplified to keep 10\% of the faces in \textit{Meshlab}~\cite{cignoni2008meshlab}.
This explicit representation of the surface is not convenient to check whether a particle is inside or outside the boundaries, and we thus employ an implicit representation stored on a uniform grid, the \ac{SDF}.
The \ac{SDF} of the capillaries was generated from the triangle mesh on a domain of size $576 \times 528 \times 24$ (simulation units).
The \ac{SDF} takes into account the periodic images of the geometry, to ensure continuity across the periodic domain, a necessary feature to perform the \ac{DPD} simulations.
Furthermore, the \ac{SDF} was merged with that of two cylinders that form the inlet and outlet of the geometry and are periodically connected (\cref{fig:sdf}).
This periodic connection makes the domain fully periodic and simplifies particle simulations as no mechanism of inserting and deleting particles is needed.

The \ac{SDF} geometric representation is usable by both \textit{Aphros}~\cite{karnakov2020aphros} and \textit{Mirheo}~\cite{alexeev2020mirheo} to perform the Stokes and the detailed blood simulations, respectively.
Furthermore it simplifies the implementation of the bounce-back of the swimmer in the simplified model, as a evaluating the \ac{SDF} at the swimmer's position indicates if it is inside or outside of the capillaries.
In the latter case, the swimmer is simply put to its previous position.

\begin{figure}
  \centering
  \includegraphics{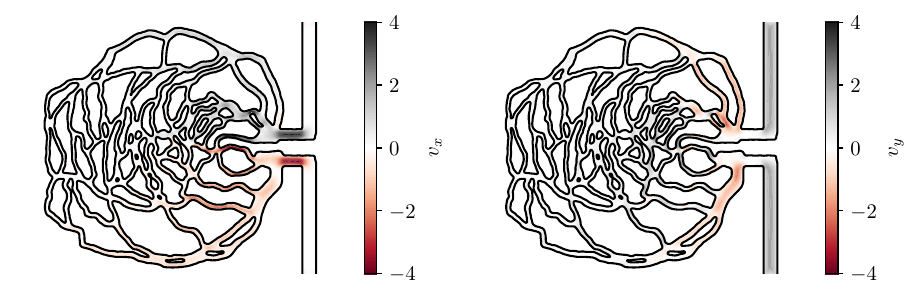}
  \caption{
    Slice of the velocity inside the capillaries obtained from the Stokes solver \textit{Aphros}.
    The left and right plots correspond to the $x$ and $y$ components, respectively, and the solid line denotes the boundaries of the walls.
  }
  \label{fig:vel}
\end{figure}

\Cref{fig:vel} shows a two-dimensional slice of the velocity field obtained from \textit{Aphros}.
This field corresponds to an inlet of unit velocity and no-slip along the walls of the capillaries, and we assumed an incompressible, Newtonian fluid.
This velocity field is used by the simplified model described in the main text.

\section{ODE model to produce line in Fig.~2 of the manuscript}

By linearity of the Stokes equation, the linear and angular velocities of a rigid object are linearly dependent on the external forces and torques applied to it~\cite{purcell1977lowRe}:
\begin{equation} \label{eq:ABF:mobility}
  \begin{bmatrix}
    \mathbf{V} \\
    \mathbf{\Omega}
  \end{bmatrix}
  =
  \begin{bmatrix}
    M^{FV} & M^{TV} \\
    M^{F\Omega} & M^{T\Omega}
  \end{bmatrix}
  \begin{bmatrix}
    \mathbf{F} \\
    \mathbf{T}
  \end{bmatrix},
\end{equation}
where the matrix coefficients are fixed in the frame of the object.
In this study there is no external force and only the external magnetic torque,
\begin{equation*}
  \mathbf{T} = \mathbf{m} \times \mathbf{B},
\end{equation*}
where $\mathbf{m}$ is the magnetic moment of the \ac{ABF} (fixed in the frame of the swimmer) and $\mathbf{B}$ is the external magnetic field.
Considering an \ac{ABF} swimming in the $x$ direction, subjected to a rotating magnetic field in the $yz$ plane,
\begin{equation}
  \mathbf{B}(t) =
  \begin{bmatrix}
    0 \\
    B \cos \omega t \\
    B \sin \omega t
  \end{bmatrix},
\end{equation}
where $\omega$ is the angular frequency of rotation of the magnetic field and $B$ its magnitude.
We assume that the swimmer rotates around the $x$ axis.
Defining $\theta$ as the angle between the $y$ axis and the magnetic moment of the swimmer, we get $\dot{\theta} = \Omega$.
The magnetic torque magnitude is given by $T_{m} = mB \sin (\omega t - \theta)$.
Using \cref{eq:ABF:mobility}, the angular velocity of the swimmer is given by $\Omega = M^{T\Omega}_{11} T_m$, hence we get the following ODE:
\begin{equation} \label{eq:ABF:ODE}
  \dot{\theta} = \omega_c \sin (\omega t - \theta),
\end{equation}
where we have defined the critical angular frequency $\omega_c = M^{T\Omega}_{11}mB$.

When $\omega \leq \omega_c$, \cref{eq:ABF:ODE} admits the steady solution $\theta(t) = \omega t + \phi$, where $\phi$ is a phase depending on the mobility matrix coefficients and the maximal magnetic torque amplitude.
In this regime, the swimmer is able to rotate at the same frequency as that of the magnetic field rotation, since the magnetic torque is large enough to compensate for the hydrodynamic torque.

On the contrary, when $\omega > \omega_c$, the magnetic torque is too small to sustain the hydrodynamic torque that would be required to keep the swimmer rotating at the same angular frequency.
The solution to \cref{eq:ABF:ODE} has been solved in refs.~\cite{schamel2013chiral,vach2013selecting} and reads
\begin{equation*}
  \theta(t) = \omega t + 2 \arctan{\left( \frac{-\omega_c + \sqrt{\omega_c^2 - \omega^2} \tan{\left(-\frac 1 2 \sqrt{\omega_c^2 - \omega^2}
    (t+\phi)\right)}}{\omega} \right)},
\end{equation*}
where $\phi$ depends on the initial conditions.
The second term in the solution is periodic in time with angular frequency $\sqrt{\omega_c^2 - \omega^2}$.
The effective (or time averaged) angular frequency of the swimmer is thus given by $\Omega = \omega - \sqrt{\omega_c^2 - \omega^2}$.

Using the mobility matrix relation, the linear velocity along $x$ is linear with the angular velocity.
Hence, we obtain that the velocity of the \ac{ABF} changes with the magnetic frequency of rotation as
\begin{equation}
  V(\omega) =
  \begin{cases}
    a \omega, & \text{if} \;\; \omega \leq \omega_c, \\
    a \left(\omega - \sqrt{\omega_c^2 - \omega^2}\right), & \text{otherwise}.
  \end{cases}
\end{equation}
The constant $a$ depends on the geometry of the swimmer, while the step out frequency depends on both the geometry and the maximum magnetic torque applied on the swimmer.

\section{Calibration of the simplified model}

\begin{figure}
  \centering
  \includegraphics{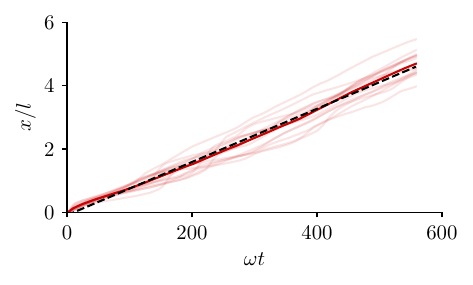}
  \includegraphics{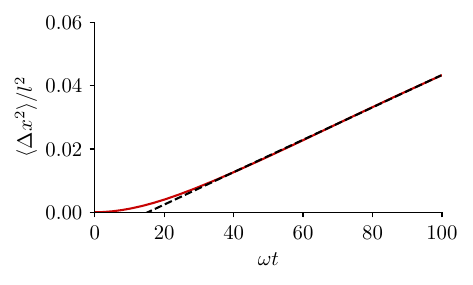}
  \caption{
    Left: Displacement $x$ of the ABF for 15 simulations (shaded lines) against time.
    The solid line corresponds to the average displacement and the dashed line is a linear fit to this line.
    Right: Mean squared deviation from the mean displacement of the ABFs over 15 simulations against time (solid line) and linear fit of the non ballistic part (dashed line).
  }
  \label{fig:diffusion:tube:x}
\end{figure}

We consider an \ac{ABF} of radius $r=\SI{2}{\micro\meter}$ and length $l=\SI{18.37}{\micro\meter}$ in a magnetic field that rotates with a frequency $f = \SI{1}{\kilo\hertz}$.
We define the angular frequency $\omega = 2\pi f$.
The swimming speed of the \ac{ABF} in blood is measured from 15 simulations with different initial cells configurations.
The simulations consist of a single \ac{ABF} swimming in a periodic tube of radius $R=\SI{10}{\micro\meter}$.
The tube is filled with a suspension of blood cells at $\Ht=25\%$ and the \ac{ABF} is initially placed at the center of the tube.
\Cref{fig:diffusion:tube:x} shows the displacement of the \ac{ABF} over time for 15 trials of the simulation, and the mean squared deviation from the mean displacement of these trials.
We estimate $U$ as the slope of the mean displacement over time, and the diffusion coefficient $D$ as the slope of $\langle \Delta x^2 \rangle$ against $6t$, where $\Delta x(t) = x(t) - Ut$.
We find a swimming speed of $U = \SI{8.40e-3}{} l \omega \approx \SI{1}{\milli\meter\per\second}$ and a diffusion coefficient (due to collisions with the surrounding cells) of $D_\mathrm{sim} = \SI{2.56e-4}{} l^2 \omega \approx \SI{542}{\micro\meter^2\per\second}$.

In the reduced order model, this corresponds to $U \approx U_{in} = 1$, where $U_{in}$ is the average velocity at the inlet and $D_\mathrm{sim} \approx 0.0123$ in simulation units.
The noise is anisotropic and to account for the differences we use a larger value during training $D=0.1 \approx 8.14 D_\mathrm{sim}$.
Furthermore, we set $U = 0.2 U_{in}$ in the simplified model, making the task more difficult but potentially more robust to the \ac{ABF} being stuck by the surrounding cells.

\section{Initial positions of the swimmers during the training phase}

\begin{figure}
  \centering
  \includegraphics{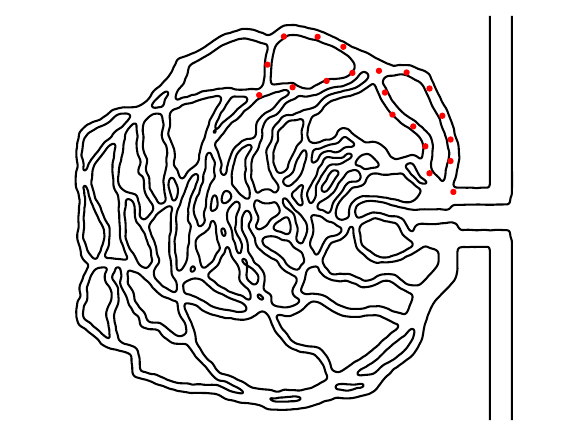}
  \caption{Position of the initial seeds to generate the initial positions of the swimmer at the beginning of each episode.}
  \label{fig:seeds}
\end{figure}

During training, the initial position of the swimmer is sampled from a Gaussian distribution with a mean $\mu$ and a standard deviation $\sigma = 8.7$.
If the sampled position is not inside the vessels or if its distance to $\mu$ is larger than $2\sigma$, a new sample is proposed, until it meets all criteria.
The mean $\mu$ is chosen as the entry of the capillary network with 20\% probability.
Otherwise, it is randomly chosen from a set of 20 positions with equal probability.
These positions were obtained as follows:
First, 400 brownian particles were initialized at the target position and evolved for a time $T_\text{max}$ with a background velocity field $-\mathbf{u}$ and the same diffusion coefficient as the swimmer.
Second, 20 points from these trajectories were sampled using the farthest point strategy~\cite{eldar1997farthest}.
These initial positions are shown in \cref{fig:seeds}.

\section{Hyper-parameters in the reinforcement learning setup}

Each episode is ended when either the \ac{ABF} reaches the target within a distance $\delta = 5$ or the simulation time is larger than $T_\mathrm{max} = 500$.
Actions are computed from the state every $\Delta t = 0.1$.
The agent was trained with a self propelling speed of $U=0.2$ and a noise level of $D=0.1$, and the maximal velocity at the inlet was set to 1.
The discount factor was set to $\gamma=1$ and we used $C=U_{in}$ in the reward equation.

We used the default settings in \textit{Korali}~\cite{martin2022korali}.
The initial exploratory noise level was set to 0.1 along each action direction.
We used the Adam optimizer with a learning rate of $\SI{1e-4}{}$ and a mini-batch size of 128.
The parameters of the policy were updated after every experience.
The replay memory has a size of 131072.
The \texttt{REFER} parameters~\cite{novati2019a} were set to $\beta=0.3$, $C_\mathrm{max}=4$, $D=0.1$, $A=0$.
The policy is represented by a neural network with 2 hidden layers of 64 units each, separated by a $\tanh$ activation function.

\section{Comparison between the DPD model and the reduced order model}

\begin{figure}
  \centering
  \includegraphics{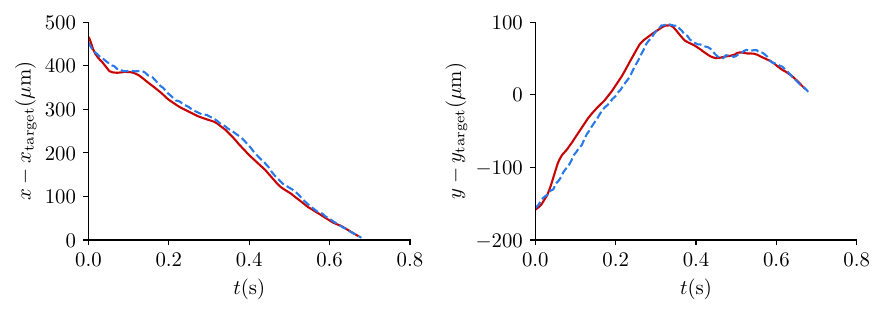}
  \caption{Comparison of the position of the swimmer with respect to its target from the DPD simulations (solid lines) and the reduced order model (dashed line) against time.}
  \label{fig:comparison}
\end{figure}

We compare the trajectories of the controlled \ac{ABF} obtained from both the \ac{DPD} simulations and the reduced order model.
These trajectories are shown in \cref{fig:comparison} and are similar with both models.
The time taken to bring the swimmer to its target is similar in both cases.

\section{Motion of the ABF in DPD simulations}

\begin{figure}
  \centering
  \includegraphics[width=0.45\textwidth]{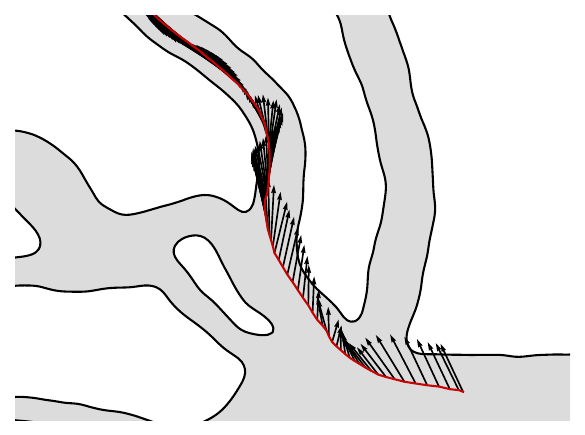}
  \includegraphics[width=0.45\textwidth]{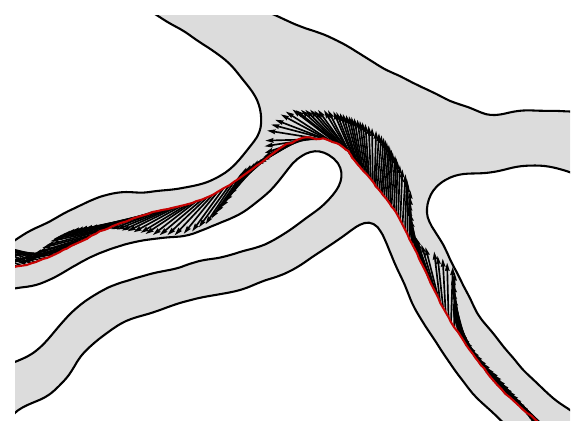}
  \caption{Trajectory of the ABF in the DPD simulation around two bifurcations (solid line) and swimming direction imposed by the policy (arrows).}
  \label{fig:dpd_traj_bifurcations}
\end{figure}

\begin{figure}
  \centering
  \includegraphics{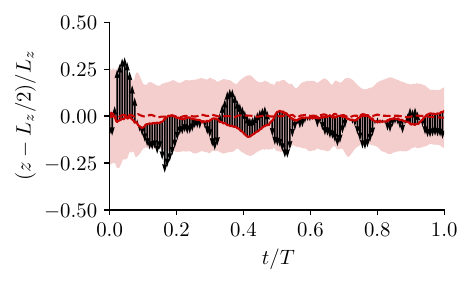}
  \includegraphics{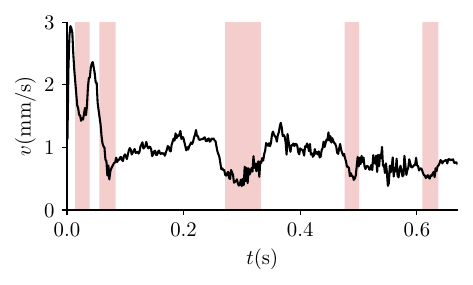}
  \caption{
    Left: $z$ coordinate of the ABF against time in the DPD simulation (solid line) and $z$ component of the swimming direction imposed by the policy (arrows).
    The shaded area indicates the wall positions in the vertical direction and the dashed line is the center of these lines.
    Right: Velocity of the ABF against time (solid line). Shaded regions indicate times when the ABF is at a bifurcation.
  }
  \label{fig:dpd_traj_z}
\end{figure}

\Cref{fig:dpd_traj_bifurcations} shows the trajectory of the \ac{ABF} in the \ac{DPD} simulation around two bifurcations.
As in the reduced order model, the trajectory tends to remain in the center of the vessels (see also \cref{fig:dpd_traj_z}) and is tangent to the imposed swimming direction when the \ac{ABF} is away from bifurcations.

However, when approaching a bifurcation, the control agent imposes a swimming direction towards the target branch, bringing the swimmer closer to the walls.
\Cref{fig:dpd_traj_z} shows that the speed of the \ac{ABF} is lower at bifurcations than in wessels away from bifurcations.
This is consistent with the controlled swimming direction not aligning with the velocity field of the brackground flow near bifurcations.

\bibliographystyle{unsrt}
\bibliography{refs}